\documentclass[twocolumn,showpacs,preprintnumbers,amsmath,amssymb]{revtex4}

\usepackage{graphicx}
\usepackage{dcolumn}
\usepackage{bm}
\begin{document}
\title{Roughness of Interfacial Crack Front: Correlated Percolation in
the Damage Zone}
\author{Jean Schmittbuhl\footnote{Permanent Address:
Departement de G{\'e}ologie, UMR CNRS 8538, Ecole Normale
Sup{\'e}rieure, 24, rue Lhomond, F--75231 Paris C{\'e}dex 05,
France. Email: schmittb@geologie.ens.fr.}
and Alex Hansen\footnote{Permanent Address: Department of Physics,
NTNU, N--7491 Trondheim, Norway. Email: Alex.Hansen@phys.ntnu.no.}}
\affiliation{International Center for Condensed Matter Physics,
Universidade de Bras{\'\i}lia, 70919--970 Bras{\'\i}lia, Distrito
Federal, Brazil}
\author{G.\ George Batrouni\footnote{Email:
george.batrouni@inln.cnrs.fr.}}  \affiliation{Institut Non-Lin\'eaire
de Nice, UMR CNRS 6618, Universit{\'e} de Nice-Sophia Antipolis, 1361
Route des Lucioles, F--06560 Valbonne, France} \date{\today}
\begin{abstract} 
We show that the roughness exponent $\zeta$ of an in-plane crack front
slowly propagating along a heterogeneous interface embeded in a
elastic body, is in full agreement with a correlated percolation
problem in a linear gradient.  We obtain $\zeta=\nu/(1+\nu)$ where
$\nu$ is the correlation length critical exponent. We develop an
elastic brittle model based on both the 3D Green function in an
elastic half-space and a discrete interface of brittle fibers and find
numerically that $\nu=1.5$. We conjecture it to be 3/2. This yields 
$\zeta=3/5$. We
also obtain by direct numerical simulations $\zeta=0.6$ in excellent
agreement with our prediction. This modelling is for the first time in
close agreement with experimental observations.
\end{abstract} 
\pacs{83.80.Ab, 62.20.Mk, 81.40.Np}
\maketitle
An important motivation for studying interfacial crack pinning
\cite{srvm95,sm97} is to simplify the study of the origin of the
scaling properties of brittle crack surfaces \cite{mpp84,bs85}.  These
scaling properties are seen, for example, in the height-height
correlation of the fracture roughness ({\it i.e.}  out-plane
roughness), which shows self-affinity.  That is, the conditional
probability density $p(x,h)$, {\it i.e.}\ the probability that the
crack surface passes within $dh$ of the height $h$ at position $x$
when it had height zero at $x=0$, shows the invariance
\begin{equation}
\label{hhscaling}
\lambda^\zeta p(\lambda x,\lambda^\zeta h)=p(x,h)\;,
\end{equation}
where $\zeta$ is the roughness exponent.  It is now generally believed
that the roughness exponent shows a universal value of about 0.80 at
larger scales \cite{blp90-mhhr92-sgr93-cw93-sss95}, while a lower
value of about 0.5 might be seen at smaller scales
\cite{dhbc96-dnbc97}.

Direct observations of the interfacial crack-pinning have been
performed recently. The problem consists of following the roughness of
a crack front moving along the flat interface between two elastically
connected blocks.  The experimental study of constrained crack
propagation between two sintered plexiglass plates presented in
\cite{sm97} resulted in the estimate of the in-plane roughness
exponent: $\zeta=0.55\pm 0.05$.  This work was followed by a longer
study leading to the estimate $\zeta=0.63\pm 0.03$ \cite{dsm99}.  A
recent study of the motion of a helium-4 meniscus along a disordered
substrate --- a problem closely related to the motion of a crack line
--- gave $\zeta=0.56\pm0.03$ \cite{prg02}.

Numerous models for interfacial crack propogation in heterogeneous
material have been proposed. The numerical simulation presented in
\cite{srvm95} is based on a perturbative Green function approach
following the quasistatic evolution of the interfacial crack front
position $a(x,t)$ \cite{gr89}, which is treated as a single-valued
function of position $x$ along the orthogonal direction to the crack
advancement direction, and time $t$. The linearized Green function
used binds together points only along the fracture front. The stress
intensity factor at a point $x$ along the fracture front is then found
to be
\begin{equation}
\label{gaorice}
\frac{K(x,t)}{K_0}=1+\frac{1}{2\pi}\ {\rm p.v.} \int_{-\infty}^{+\infty}
\frac{a(x',t)-a(x,t)}{(x-x')^2}\ dx'\;,
\end{equation}
where $K_0$ is the stress intensity factor that would result if the
crack were straight \cite{gr89}.  The fracture is advanced by
identifying the most stressed point along the fracture line and
advancing this by a small step.  The roughness exponent of the crack
front was estimated numerically to be $\zeta=0.35$, while a direct
dynamical renormalization group calculation gave $\zeta=1/3$ to lowest
order \cite{ek94}.  Higher-order corrections to this quasistatic
analysis increases the value of $\zeta$ to 0.48 \cite{cdw01}, while a
different quasistatic analytical technique suggests $0.390\pm0.002$
\cite{rk02}. Dynamic effects have beeen largely studied numerically
and analytically \cite{pr94-rf97-rf98-mr98-mr00}, in particular in the
form of crack front waves. They lead to an increased roughness
exponent compared to the initial 1/3-value up to $\zeta=0.5$ but still
smaller than experimental observations.

Hence, the situation today is that there is a large and significant
discrepancy between theoretical and experimental estimates: theoretical
estimates being considerably smaller than the experimental ones.

\begin{figure}
\includegraphics[width=2.7cm,angle=270]{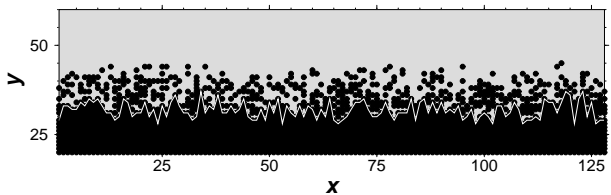}
\caption{The crack front for a $128\times 128$ system. The fracture is
propagating from bottom to top.  The broken springs are black dots. The 
crack front is drawn as a white line.}
\label{fig1}
\end{figure}

In this letter we present a numerical calculation of the roughness
exponent of a crack front that propagates quasi-statically along a
heterogeneous interface, based on a Green function technique which
differs from that previously used. Indeed it does not reduce the crack
tip to a single tortuous line but describes the tip as a region of
interactions between microcracks (see Fig.~\ref{fig1}). Our technique
is based on the static solution of the elastic equations for the
deformation of the surface of an elastic half space \cite{ll58}.  The
local deformation, $u$, at position $(x,y)$ along the plane is related
to the normal stress field $\sigma$ by the expression
\begin{equation}
\label{gruf}
u(x,y)=\int\int G(x-x',y-y') \sigma(x',y')\ dx'\ dy'\;,
\end{equation}
where the Green function is \cite{j85}:
\begin{equation}
G(x,y)=\frac{1-s^2}{\pi e }\
\frac{1}{|(x,y)|}\;,
\label{thanksjohn}
\end{equation}
with $s$ is the Poisson ratio and $e$ the elastic constant.  For large 
$r=|(x-x',y-y')|$, $G\sim1/r$.
For the sake of simplicity, as generally done for the study of contact
between two elastic bodies \cite{j85}, we substitute one of the
elastic plates by an infinitely rigid one.  The other plate is modeled
as an elastic block for which Eq.\ (\ref{thanksjohn}) is valid.  We
discretize the model, so that the forces and deformations are
described by the discrete version of Eq.\ (\ref{gruf}),
\begin{equation}
\label{M1}
u_i=\sum_j \overline{G}_{i,j} f_j\;.
\end{equation}
$\overline{G}_{i,j}$ is the Green function  Eq.\ (\ref{thanksjohn})
averaged over an area $b^2$,
\begin{equation}
 \overline{G}_{i,j}=\frac{1}{b^2}\ \int_{-b/2}^{+b/2} \int_{-b/2}^{+b/2}\ dx'\
dy' G(i-x',j-y')\;,
\label{avethanksjohn}
\end{equation}
where $b$ is the lattice spacing,  $u_i$ is the deformation of the
elastic body at site $i$, and $f_i$ the force acting at that point.
The indices $i$ and $j$ in Eq.\ \ref{M1} run over all $L^2$ sites.

The elastic block is connected to the infinitely stiff plate by a
discrete interface made of an array of elastic harmonic springs.  Each
spring is brittle and has a breaking threshold randomly drawn from a
uniform distribution between zero and one.  The spacing between the
springs is $b$ in both the $x$ and $y$ directions.  The force $f_i$
that an unbroken spring $i$ is carrying, is transferred over an area
of size $b^2$ to the soft surface and given by Hooke's law: 
\begin{equation}
\label{hooke}
f_i = -k(u_i-D)\;,
\end{equation}
where $k$ is the spring constant ($k=1$ for all springs). $D$ is the
displacement of the infinitely stiff medium, and is a function of $y$,
{\it i.e.} solid rotation of the stiff medium.  The quantity $(u_i-D)$
is, therefore, the length that spring $i$ is stretched.  We assume
periodic boundary conditions both in the $x$ and $y$ directions.  In
order to model the mode~I fracturing of the interface between the two
media (Double Cantilever Beam (DCB) test) in a way compatible with the
biperiodic boundary conditions, we let $D(y)$ be wedgeshaped with a
maximum at $y=0$ and $L$, and zero for $y=L/2$ ({\it i.e.} mirroring
conditions).  The load the system, $D(y)$ is increased and the springs
break one by one.  The numerical technique to solve the equations that
ensue is presented in detail in \cite{bhs02}.

\begin{figure}
\includegraphics[width=6.5cm,angle=270]{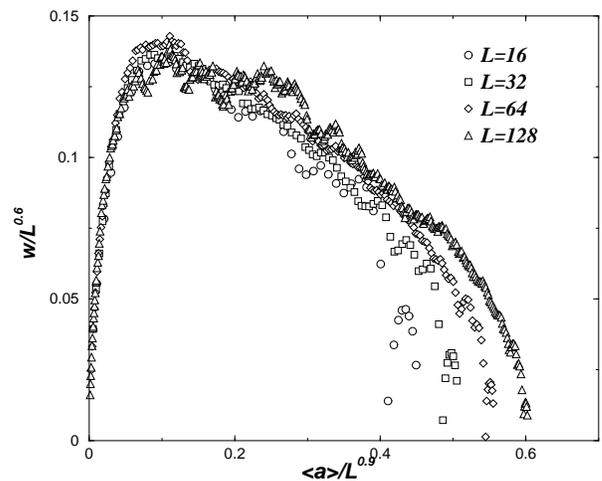}
\caption{Family-Vicsek scaling of the crack front roughness:
$w/L^{0.6}$ vs.\ $\langle a\rangle /L^{0.9}$ for different system
sizes: $L=16$, $32$, $64$ and $128$.}
\label{fig2}
\end{figure}

We show in Fig.~\ref{fig1} a typical damage front in a $128\times 128$
system.  We define the fracture front $a(x,t)$ in this model as the
set of nodes that form a continuous path separating the infinite
cluster of broken springs from the infinite cluster of unbroken
springs. This definition is similar to that of Ref.~\cite{sm97}. From
Fig.~\ref{fig1}, we clearly see that even if a front can be defined, an
extended damage zone exists. Accordingly the front does not capture
all the active tip of the fracture. This observation supports the
relevance of a non line description of the crack front.

In Fig.~\ref{fig2}, we show the width of the fracture front $w =
\sqrt{\langle a^2\rangle -\langle a\rangle^2}$ as a function of its
average position $\langle a\rangle$ --- which acts as a time parameter
in this quasi-static model, for various system sizes.  By collapsing
the width evolution for the different system sizes, we see that the
crack front follows a Family-Vicsek scaling \cite{fv85}. Two important
exponents are, thus, obtained: the roughness exponent
$\zeta=0.6$ and dynamical exponent $z=0.9$. The roughness exponent is
in close agreement with the experimentally obtained value
\cite{sm97,dsm99,ms01,prg02}, while the dynamical exponent $z$ was
found in Ref.~\cite{ms01} to be $1.2$.

Our model distinguishes itself from earlier numerical models in three
major ways: (1) Most previous models are based on a small perturbation
approach and include linearizations \cite{gr89} that are not used
here.  Furthermore, it was assumed in the earlier studies that (2) the
fracture line was a single-valued function, hence ruling out
overhangs.  This assumption prevents islands of unbroken bonds from
forming in the wake of the advancing fracture line.  (3) The
assumption of a single advancing fracture line also prevents the
formation of a damage zone in front of the fracture line.  None of
these three assumptions are necessary in the present model.  In order
to test whether assumptions (2) or (3) are responsible for the
difference in roughness exponent, we imposed both on the present
model.  No change in $\zeta$ was observed.  Hence, we conclude that it
is the linearization assumption that is responsable for the
difference.

\begin{figure}
\includegraphics[width=6.5cm,angle=270]{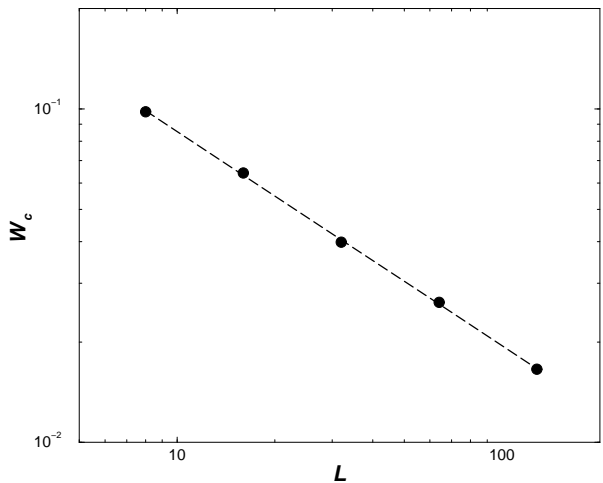}
\caption{Fluctuations of the density of broken bonds $W_c=(\langle
p_c^2\rangle -\langle p_c\rangle^2)^{1/2}$ plotted against $L$ for a
homogeneous parallel loading without any gradient, {\it i.e.} constant
D over the whole system.  The slope of the straight line is
$-1/\nu=-0.65$.}
\label{fig3}
\end{figure}

In the following, we demonstrate that the roughness exponent $\zeta$
is a result of a correlated percolation process in a gradient imposed
by the mode I loading of the system.  We start by charactarizing the
correlated percolation. For this goal, we consider a similar problem
but where the loading is obtained without any gradient, {\it i.e.}
horizontal and parallel displacement of the rigid medium
\cite{bhs02}. In this case, when the homogeneous displacement $D$ of
the rigid medium reaches a maximum value, $D_c$ the system goes
unstable, and unless $D$ is decreased again, catastrophic failure sets
in.  In Ref.~\cite{bhs02}, the size distribution of clusters of broken
springs was studied, and a power law was found with an exponent
$\tau=1.6$.  This value is different from ordinary percolation where
$\tau \approx 2.05$ \cite{sa92} and shows that correlated percolation
takes place.  In Fig.~\ref{fig3}, we show the fluctuations of density
of broken springs at $D=D_c$.  If there is a diverging correlation
length in the problem, these fluctuations scale as $L^{-1/\nu}$.  We
find that $1/\nu=0.65$, leading to $\nu=1.54$.  We conjecture that the
exact value is $\nu=3/2$.  Hence, the fracture process in this system
is in a universality class which is {\it different\/} from standard
percolation where $\nu =4/3$.

\begin{figure}
\includegraphics[width=6.5cm,angle=270]{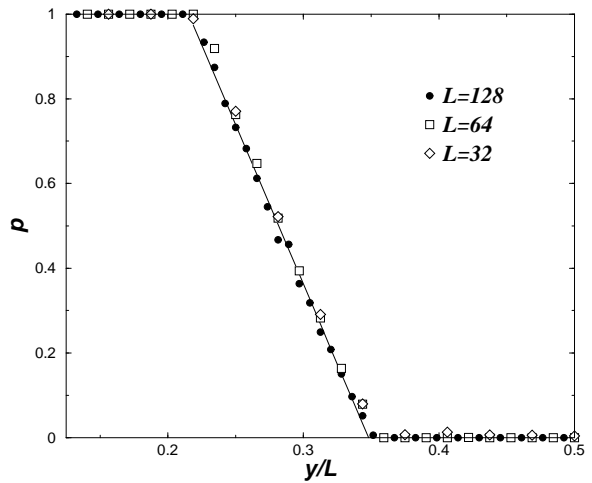}
\caption{Damage profile $p(y/L)$ in reduced variable $y/L$ for three
different system sizes $L$.}
\label{fig4}
\end{figure}

When $D$ is no longer constant, but is given by the wedge shape (DCB
load), the form of interactions between the springs does not change.
Hence, the critical properties of the model when run in the
``horizontal" mode are still present under DCB loading.  The wedge
shape of $D$ leads to a gradient in the loading of the springs, going
from very high loading where the damage is large to very low loading
well into the still intact part of the interface.  In Fig.~\ref{fig4},
we show the damage profile $p(y)$, {\it i.e.} the density of broken
springs averaged in the $x$ direction, for systems of different sizes.
The profile is clearly linear.  Somewhere along this damage profile,
there is a line in the $x$-direction at $y=y_c$ where the damage
density is critical, $p(y_c)=p_c$.  In the vicinity of this line,
there is a critical region which is characterized by being on the edge
between stability and instability and corresponds to the crack
front. Following the arguments of Sapoval {\it et al.\/} for
percolation in a gradient \cite{srg85}, if $p(y)$ follows a linear
law,
\begin{equation}
\label{linearp}
p(y)=1-\frac{y-y_0}{l_y}\;,
\end{equation}
on the interval $y_0 \le y \le y_0+l_y$, where $l_y$ is the length scale
characterizing the width of the damage zone, then $y_c=y_0+l_y(1-p_c)$
is the position of the fracture front (when ignoring small corrections for
finite-size systems).  The critical region has a width $\xi=|y_w-y_c|$
which is related to the damage $p(y)$ as $\xi \sim |p(y_w)-p_c|^{-\nu}$.
Eliminating $|y_c-y_w|$ between these two expressions for $\xi$ gives
\begin{equation}
\xi\sim l_y^{\frac{\nu}{1+\nu}}\;.
\end{equation}

Since the gradient in $D$ is inversely proportional to $L$ and the
average strength of the springs does not change with $L$, we have that
\begin{equation}
\label{lyL}
l_y \propto L\;.
\end{equation}
In Fig.\ \ref{fig4}, the reduced variable $y/L$ was used resulting in
data collapse for different system sizes, thus validating Eq.\
(\ref{lyL}).  Furthermore, the width of the fracture front, $w$ is
proportional to the width of the critical region $\xi\propto w$.  Hence, we
find
\begin{equation}
\label{finally}
w\sim L^{\frac{\nu}{1+\nu}} = L^{3/5}\;.
\end{equation}
where we have used our estimate $\nu=3/2$.  This result is in
excellent agreement with our numerical simulations and with the
experimental results. 

A similar idea has been proposed for the outplane roughness of
fracture surfaces where the gradient and the $\nu$ exponent are
different and leads to $\zeta=4/5$ in excellent agreement with the
experimental value 0.80 \cite{hs02}.  We finally note that Zapperi
{\it et al.\/} \cite{zhr00} have pointed out the connection between
gradient percolation and the interface fracture problem in the limit
of infinitely stiff plates.  We have shown in this letter that this
mechanism stays intact also when the plates do respond elastically.
However, the universality class of the corresponding correlated
percolation problem is different from ordinary percolation, leading to
the observed roughness exponent $\zeta=3/5$ rather than 4/7 for
ordinary percolation in a gradient.

This work was partially funded by the CNRS PICS contract $\#753$ 
and the Norwegian Research Council, NFR. J.S.\ and A.H.\  thank 
Fernando A.\ Oliveira
and the ICCMP for financial support during our stay in Bras{\'\i}lia.
Discussions with K.\ J.\ M{\aa}l{\o}y are gratefully acknowledged.

\end{document}